# First high peak and average power single-pass THz FEL based on high brightness photoinjector

M. Krasilnikov†, Z. Aboulbanine[1], G. Adhikari[2], N. Aftab, A. Asoyan[3], P. Boonpornprasert, H. Davtyan[3], G. Georgiev, J. Good, A. Grebinyk, M. Gross, A. Hoffmann, E. Kongmon[4], X.-K. Li, A. Lueangaramwong[5], D. Melkumyan, S. Mohanty, R. Niemczyk[6], A. Oppelt, H. Qian[7], C. Richard, F. Stephan, G. Vashchenko, T. Weilbach[8], X. Zhang[9], Deutsches Elektronen-Synchrotron DESY, Zeuthen 15738, Germany

M. Tischer, E. Schneidmiller, P. Vagin, M. Yurkov, E. Zapolnova, Deutsches Elektronen-Synchrotron DESY, Hamburg 22607, Germany

W. Hillert, J. Rossbach, University of Hamburg, Hamburg 22761, Germany

A. Brachmann, N. Holtkamp, H.-D. Nuhn, SLAC, Menlo Park, CA 94025, USA

[1]now at Oak Ridge National Laboratory, USA
[2]now at SLAC National Accelerator Laboratory, USA
[3]on leave from CANDLE Synchrotron Research Institute, Armenia
[4]on leave from Chiang Mai University, Thailand
[5]now at Diamond Light Source Ltd, UK
[6]now at Helmholtz-Zentrum Dresden Rossendorf, Germany
[7]now at Zhangjiang Lab, China
[8]now at Paul Scherrer Institute, Switzerland
[9]on leave from Tsinghua University, China

## Abstract

Advanced experiments using THz pump and X-ray probe pulses at modern free-electron lasers (FELs) like the European X-ray FEL require a frequency-tunable (from 0.1 THz to 30 THz), high-power (>10 microjoule), narrow-band (~1-2%) THz source maintaining the repetition rate and pulse structure of the X-ray pulses. This paper reports the first results from a THz source, that is based on a single-pass high-gain THz FEL operating with a central wavelength of 100 micrometers. The THz FEL prototype is currently in operation at the Photo Injector Test facility at DESY in Zeuthen (PITZ) and uses the same type of electron source as the European XFEL photo injector. A self-amplified spontaneous emission (SASE) FEL was envisioned as the main mechanism for generating the THz pulses. Although the THz FEL at PITZ is supposed to use the same mechanism as at X-ray facilities, it cannot be considered as a simple scaling of the radiation wavelength because there is a large difference in the number of electrons per radiation wavelength, which is five orders of magnitude higher for the THz case. The bunching factor arising from the electron beam current profile contributes strongly to the initial spontaneous emission starting the FEL process. Proof-of-principle experiments were done at PITZ using an LCLS-I undulator to generate the first high-power, high-repetition-rate single-pass THz FEL radiation. Electron bunches with a beam energy of ~17 MeV and a bunch charge of up to several nC are used to generate THz pulses with a pulse energy of several tens of microjoules. For example, for an electron beam with a charge of ~2.4 nC, more than 100 microjoules were generated at a central wavelength of 100 micrometers. The narrowband spectrum was also demonstrated by spectral measurements. These proof-of-principle experiments pave the way for a tunable, high-repetition-rate THz source providing pulses with energies in the millijoule range.

## Introduction – Requirements for THz source

Intense, short X-ray pulses from modern light sources based on the free-electron laser (FEL) concept enable imaging on the atomic scale. A typical method to generate high power X-ray pulses is the so-called self-amplified spontaneous emission free-electron laser (SASE FEL) [1], [2], [3], [4], [5]. In a SASE FEL the radiation from the electron shot noise is amplified in a long undulator to produce coherent X-rays of extremely high brightness. Pump–probe experiments at XFEL facilities are a powerful tool for exploring the interaction between matter and light at ultrafast time scales and atomic resolution. The combination of such X-ray probe pulses with infrared/terahertz (IR/THz) pump pulses opens up a wide range of new scientific opportunities, because many excitation mechanisms of matter have resonances in the THz range. Advances in the development of intense THz sources over the past decade [6], [7], [8], [9] have revealed the demand for the capability to perform these time-resolved THz/X-ray experiments at various XFEL facilities [10], [11], [12], [13], [14]. For high efficiency pump-probe experiments, it is highly desirable to accompany each X-ray probe pulse with a THz pump pulse to maintain the same pulse train time structure, which requires a MHz repetition rate of THz pulses within the pulse train at a repetition rate of tens of Hz. There is a shortage of THz radiation sources with high average and peak power that are tunable within the THz gap in the electromagnetic spectrum and capable of supporting the flexibility of the pulse train structure of XFEL facilities. The THz source requirements for the European XFEL [9] are a THz frequency tunability from 0.1 THz to 30 THz, while the most challenging range is 3 THz to 20 THz. A minimum THz pulse energy of 10 μJ should be achievable at all frequencies. In addition, it must match the European XFEL pulse time structure of pulse trains up to currently 600 μs length at a repetition rate of 10 Hz, and a maximum pulse frequency in a pulse train of 4.5 MHz – up to 27000 pulses per second.

There are two classes of THz sources which could be considered for pump-probe experiments at XFELs: conventional laser-based sources and accelerator-based sources. The first class usually uses optical rectification with $LiNbO_3$ or organic crystals and is known to be limited in frequency tunability (0.1-0.6 THz). It is still challenging to achieve a MHz repetition rate with these sources [9], [15]. The accelerator-based THz sources provide more flexibility in tunability and pulse train structure. Several mechanisms can be utilized to generate THz radiation from accelerated electron beams. Wide bandwidth sources are realized using coherent transition or diffraction radiation (CTR or CDR) as well as radiation from bending magnets [16]. To achieve high energies of THz pulses from these sources, short electron bunches are required to maintain longitudinal coherence of the radiation. Therefore, for THz radiation at vacuum ultraviolet (VUV) and XFEL facilities, "afterburning" is often considered, which uses short beams that had already radiated at shorter wavelengths [17], [8], [18], [19]. A relatively long distance between beam dump and XFEL user stations makes the THz radiation transport over hundreds of meters a challenge for this option [20], [21].

Short electron bunches with bunch length shorter than the THz radiation wavelength are also required for narrow-band sources using undulators, the so-called superradiant sources [22], [23], [24]. A single-pass FEL using short 200 pC, 5.5 MeV electron bunches to radiate at a resonant frequency of 160 GHz from a helical undulator with zero-slippage demonstrated high efficiency (~10%) [25].

Low gain THz FEL oscillators are usually used to generate THz radiation with high temporal and transverse quality [26], [27], [28]. With a reasonable length of the optical resonator used in such sources, a rather high repetition rate of electron microbeams is required, which turns out to be relatively low brightness of the electron beam, in particular, low bunch charge. With available undulators and moderate beam energies, the THz pulse energies are up to now limited by the attainable bunch charges that can be used in sub-ps electron bunches.

The short bunch limitation can be overcome by making use of a SASE FEL where the coherent radiation is generated by electron bunches much longer than the radiation wavelength $\lambda_{rad}$ due to the development of microbunching at this wavelength [29]. Earlier developments using this mechanism [30] reported powerful (~18 MW) radiation at 640 μm using a 2 MeV, ~1 kA electron beam, which was considered a significant advance in power for sources available for sub-THz frequencies.

Developments of a prototype for a high-power tunable accelerator-based THz source for pump-probe experiments at the European XFEL are ongoing at the Photo Injector Test Facility at DESY in Zeuthen (PITZ). The primary goal of PITZ is to develop high brightness electron sources for the European XFEL photo injector [31], [32]. The existing infrastructure with a high-brightness electron source makes PITZ an ideal site for the development of an accelerator-based THz source because it can generate the high charge beams (up to several nC) required for high-power THz radiation and it has a pulse train structure identical to the one of the European XFEL. The available electron beam parameter space at PITZ was used to investigate the expected THz performance of the SASE FEL assuming suitable undulator parameters. The wavelength of undulator radiation $\lambda_{rad}$ emitted by an electron beam with Lorentz factor $\gamma$ from an undulator with period $\lambda_U$ is determined by the formula

$$\lambda_{rad} = \frac{\lambda_U}{2\gamma^2}(1 + K^2 + \gamma^2\theta^2), \qquad (1)$$

where $\theta$ is the angle of observation. The helical undulator parameter is $K = \frac{\lambda_U e H_U}{2\pi mc^2}$, where $H_U$ is the peak magnetic field, $e$ and $m$ are the charge and mass of electron, and $c$ is the speed of light. For a planar undulator, $K/\sqrt{2}$ should be used instead of $K$ in Eq. (1).

An APPLE-II[1] type helical undulator with 40 mm period and undulator parameter $K$=1.8 was considered for rough modeling of lasing at a central wavelength of 100 μm for a beam energy of ~15 MeV. It was found that the peak current of the electron beam is the most influential parameter for high energy THz radiation pulses. A bunch charge of 1-6 nC (experimentally demonstrated at PITZ) was assumed for electron beams with an RMS duration of ~10 ps, giving a peak current in the range 40 A-240 A. Using the FEL theory [29], [33] for these parameters, one can obtain high THz pulse energies (at the mJ level) with a reasonable FEL instability gain $\Gamma$ (see equation (5) in *Appendix D*), the inverse gain factor is plotted in Figure 1a. The calculated FEL parameter $\rho$~0.008-0.016 corresponds to a rather high efficiency (~1.5%) of the modelled THz SASE FEL with a bandwidth of ~2-4%.

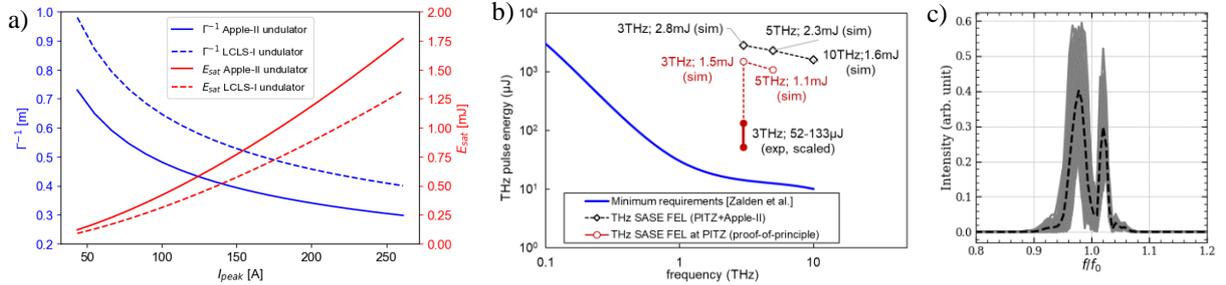

Figure 1: a) Preliminary modeling of a THz SASE FEL at a central wavelength of 100 μm using the PITZ electron beam parameters and using two different undulator types: an APPLE-II helical undulator (solid curves) and an LCLS-I undulator (dashed curves). Inverse FEL gain (left axis) and THz pulse energy estimated at the saturation point (right axis) as a function of the electron beam peak current are shown. b) THz pulse energy as a function of the central radiation frequency. Blue curve shows the minimum pulse energy required for user experiments [9]. Filled markers depict pulse energies experimentally demonstrated at PITZ with and without band-pass filter (filled red circles for ~30 μJ and 65 μJ measured, correspondingly, scaled up with the THz transport loss factors to 52 μJ and 133 μJ) using 2-2.4 nC Gaussian beams. Pulse energies expected from simulations for a PITZ-like based THz SASE FEL with APPLE-II (empty black rhombi) and current PITZ setup with LCLS-I undulator (empty red circles) using 4 nC flattop beams are shown as well. c) Simulated spectrum for the current PITZ setup with the LCLS-I undulator for the 4 nC flattop beam case, $f_0 = 3$ THz.

The predicted performance estimates for the THz SASE FEL prompted for more detailed simulations and experimental studies. It was decided to perform a proof-of-principle experiment on the single-pass high-gain THz FEL at PITZ using an LCLS-I undulator, which was available on loan from SLAC [34]. Rough modeling of the expected performance of the proof-of-principle experiment is shown in Figure 1a. Results of more detailed (start-to-end) simulations of the PITZ setup with the LCLS-I undulator for radiating at 100 μm and 60 μm central wavelength by using 4 nC electron beams are shown in Figure 1b, together with achieved experimental data. The simulated spectrum for the 100 μm case is shown in Figure 1c. To illustrate the perspectives of the approach the results of the start-to-end simulations for the PITZ-like setup with optimized APPLE-II undulator are shown as well. It should be noted that all these simulations were performed using the GENESIS code [35], which is not capable of modeling waveguide effects. Despite this idealization, the simulations give a good estimate of the expected energy level of the THz radiation pulse.

---
[1] Advanced Planar Polarized Light Emitting – II, for more details see e.g. [66]

## Photo Injector and THz beamline

The PITZ accelerator (Figure 2) consists of a radiofrequency (RF) photogun and an RF booster cavity, both standing wave resonators operating at 1.3 GHz and powered by two separate 10 MW klystrons. Currently, a new generation of normal conducting L-band electron gun (Gun5) is in operation at PITZ [36]. Gun5 is capable of operating with an RF pulse length of up to 1 ms at 10 Hz repetition rate. The photocathode laser is capable of providing pulse frequency up to 4.5 MHz in a pulse train (up to 4500 pulses per train) repeated at 10 Hz, potentially increasing the maximum number of electron bunches per second to 45000. The RF gun with an accelerating gradient of ~60 MV/m and a high quantum efficiency (QE~5-10%) $Cs_2Te$ photocathode is capable of generating high-charge (up to several nC) electron bunches with a maximum beam momentum of ~6.67 MeV/c. This photocathode technology has proven to be reliable, with a lifetime of months to years [37]. Further beam acceleration is realized utilizing the booster cavity reaching a final beam momentum of up to ~24 MeV/c. The PITZ accelerator incorporates comprehensive electron beam manipulation and diagnostics, including several sets of quadrupole magnets, scintillator screen and beam emittance measurement systems, beam position monitors, and charge measurement instruments.

In order to enable the THz proof-of-principle experiment, the accelerator beamline had been modified and extended into a second tunnel (tunnel annex) where the THz beamline was installed [38]. Figure 2 shows the components related to the production, transport and matching of electron beams for THz generation. To extend the range of operation modes a bunch compression chicane was added to the beamline in the main PITZ tunnel, but has not yet been used for THz generation studies. One of the space constraints of the current setup is the 1.5-meter concrete wall between the two tunnels, through which only the beam tube passes and no focusing magnets, which is a limiting factor for beam transport and matching in the tunnel extension. The THz beamline in the second tunnel includes a matching section, the LCLS-I undulator (see

*Appendix A* for some details on the undulator module used) centered at $z_{center} = 29.9\ m$ from the cathode, followed by the THz diagnostic section (see [39] and *Appendix C*). Three THz diagnostics stations are shown in Figure 2 (TD1 – upstream of the undulator, TD2,3 – downstream of the undulator), each associated with screen stations (High3.Scr1-3) for transverse electron beam diagnostics.

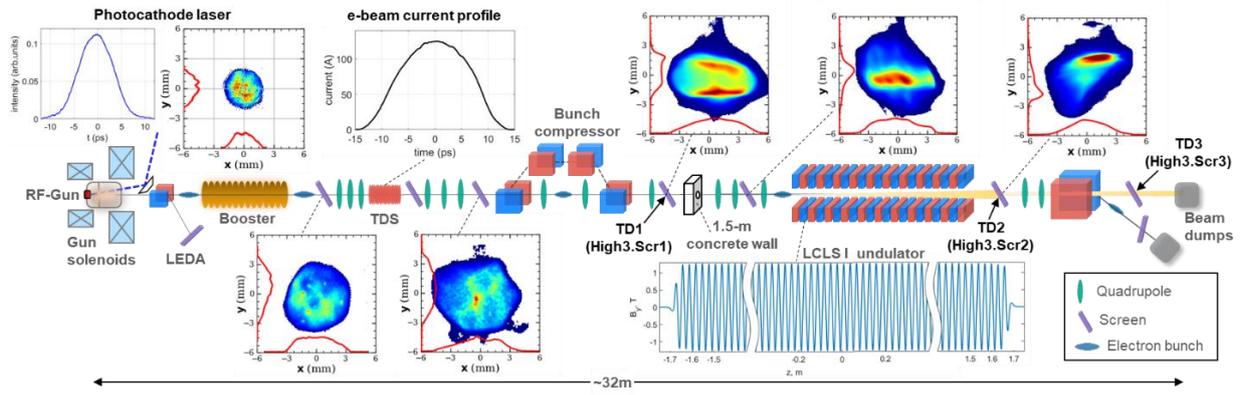

Figure 2: PITZ machine layout (not to scale). The beam travels from the RF gun through the booster, electron beam transport beamline, bunch compressor, the LCLS-I undulator, and THz diagnostic section. The photocathode laser and electron beam transverse distributions and temporal profiles measured along the beamline are shown as insets close to locations of measurement. The measured magnetic field of the LCLS-I undulator is shown as well.

## THz SASE FEL lasing

The THz beamline for the proof-of-principle experiments with an LCLS-I undulator at PITZ was completed in summer 2022 followed by commissioning. The challenge of matching high charge electron beams into the undulator was successfully accomplished, and the first lasing at a central wavelength of 100 μm was achieved in August 2022 [40]. A special procedure for accurate matching of the strongly space charge dominated electron beams into the LCLS-I undulator has been developed and applied (see *Appendix B*). Further optimization of the pyrodetector signal at TD3 (see Figure 2) with band-pass filter resulted in measured FEL gain curves that showed clear saturation of the THz pulse energy $W$ at a level of tens of μJ. The measured mean energy of THz pulses $\langle W \rangle$ using 500-shots statistics for different active undulator lengths is shown in Figure 3 for two values of the bunch charge. Before measuring the gain curve, the THz pulse energy was optimized using the entire undulator length separately for each charge. The relative RMS fluctuation rates ($\sigma_W/\langle W \rangle$) are plotted on the right axis in Figure 3. It should be noted that the first low energy points for the gain curves were measured at the same gain of the pyro detector amplifier as the high energy points, and the accuracy suffered because of the poor signal-to-noise ratio. The measured curves plotted on a log scale show a strong dependency of the THz pulse energy on the bunch charge (21.4 μJ for 2 nC versus 6.1 μJ for 1 nC). This is a clear sign of coherent FEL lasing. The backward propagation of the exponential range of the gain curves to the undulator entrance (z=0) leads to initial signals at a pJ level, which is in basic agreement with the expected noise level at this wavelength. The estimated FEL gain of ~$10^6$ indicates a high-gain THz FEL, which agrees well with theoretical expectations. Furthermore, the experimentally detected maximum energy of the THz pulse already exceeds the above-mentioned minimum user requirements.

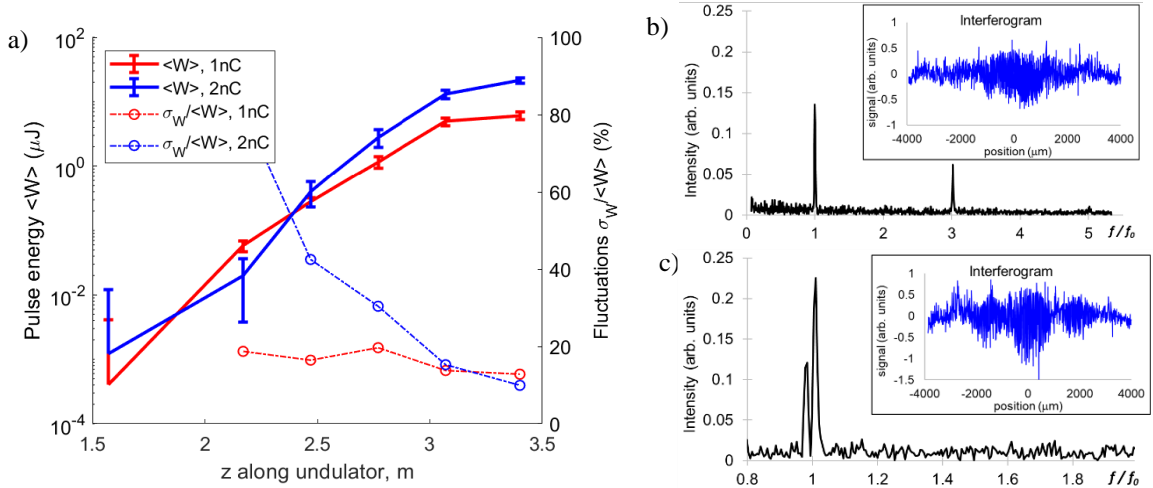

Figure 3: a) Gain curves for 1 nC and 2 nC electron bunches measured at TD3 (see Figure 2) using the band-pass filter. The corresponding pulse energy fluctuation rates are plotted on the right axis. The maximum pulse energy measured: $\langle W \rangle$@1 nC = (6.1 ± 0.8) μJ and $\langle W \rangle$@2 nC = (21.4 ± 2.1) μJ. Radiation spectrum measured with spectrometer at the end (b) and in the middle (c) of the gain curve. The measured fundamental frequency $f_0$ is 2.82 THz. Insets are the corresponding interferograms.

The first spectral measurements were performed at this screen station using a compact broadband THz spectrometer based on the reflective lamellar grating [23]. The band-pass filter was not used for these measurements. The spectrum measured for the full length of the undulator is shown in Figure 3b. A narrow-band spectrum is centered at 2.82 THz ($\lambda_{rad} \approx 106.6$ μm), which agrees well with the expected on-axis wavelength of $\approx 100.2$ μm (mean electron beam momentum of ~16.6 MeV/c) when considering the collected radiation for the observation angle of ~21 mrad (see formula (1)), which is consistent with the geometry of the TD3. The measured bandwidth of ~1.7% matches well with the expected theoretical value. It is noteworthy that higher odd harmonics were also detected, the third harmonic is well expressed in the Figure 3b, the signature of the fifth harmonic can also be recognized. The spectrum measured with ~65% of the undulator length (2.4 m of 3.4 m) corresponding to the linear regime of the FEL instability is shown in Figure 3c. The two-peak structure is similar to that obtained from the simulations (see Figure 1c and discussions on modeling below).

The strong domination of space charge effects in the narrow vacuum chamber of the undulator makes optimization of the radiation pulse energy not straightforward: slight changes to the trajectory and beam envelope matching conditions at the undulator entrance cause different electron beam halo losses on the vacuum chamber walls and change the scenarios for the overlap of radiation field and particles. In addition, the radiation pulse energy dependence on the bunch charge is suppressed because the longitudinal space charge forces increase the bunch length and prevent the peak beam current from scaling linearly with the bunch charge.

Optimization of the THz FEL radiation was also performed at the TD2 station (Figure 2), located ~0.5 m downstream of the undulator exit. The station uses an in-vacuum mirror with a 5 mm diameter hole through which the electron beam passes, and is not equipped with a bandpass filter. Such a setup without band-pass filter implies, on the one hand, the possibility of detecting higher order odd harmonics (third and fifth) over the main wavelength of 100 μm. A very rough estimate based on a linear model gives a contribution of higher-order harmonics of less than 10% to the total generated radiation. On the other hand, the radiation loss through the hole in the mirror is much higher for higher harmonics, and their detection ratio at TD2 is therefore significantly lower. As a starting point for the optimization at TD2, 2 nC electron beam transport and matching parameters close to those obtained from the optimization at TD3 were used (blue curves in Figure 3). A maximum THz pulse energy of ~15 μJ was measured at TD2 without band-pass filter for 2 nC bunch charge. Simultaneous measurements at TD3 of the radiation that passed through the hole in the vacuum mirror at TD2 yielded ~5 μJ, whereas transporting the full beam to TD3 (with the mirror at TD2 removed) resulted in ~22 µJ. The next measurement campaign was aimed at maximizing the THz pulse energy on High3.Scr2 for a bunch charge increased to 2.4 nC. A multivariate Bayesian optimizer involving electron beam trajectory, envelope, mean energy, and energy chirp was used [41]. As a result, a record high energy of the registered pulse of ~65 µJ was achieved. Considering the radiation collection efficiency (~50% for TD2 as described in *Appendix C*), radiation pulses with more than 100 µJ must have been generated. Results of these optimizations are shown in Figure 4a. The obtained gain curve clearly shows a signature of the saturation (Figure 4b). The pulse energy fluctuation rate is reduced to 5.5-6.5% (RMS) towards the end of the undulator.

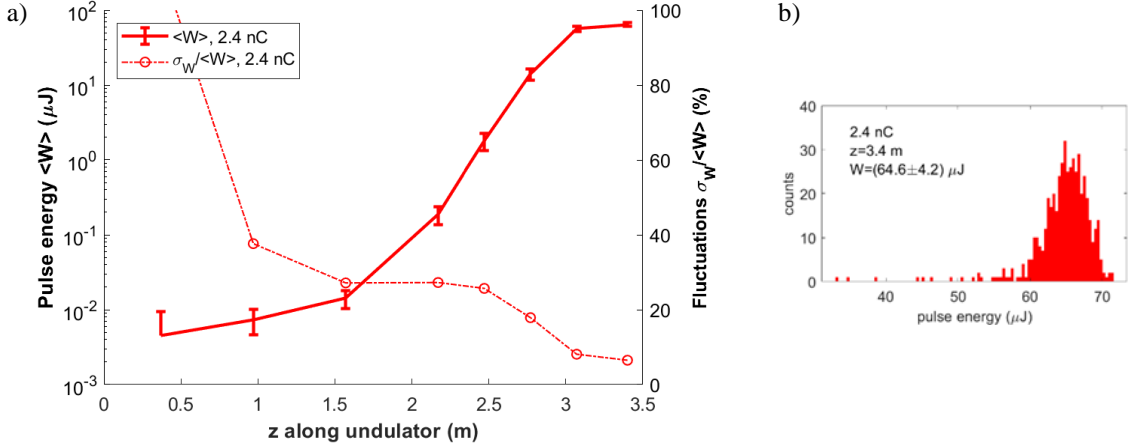

Figure 4: a) Gain curve measured at TD2 without band-pass filter for bunch charges of 2.4 nC. The corresponding fluctuation rates are plotted on the right axis. b) THz radiation pulse energy statistics (500 shots) for the highest measured mean pulse energy $\langle W \rangle @2.4\ \text{nC} = (64.6 \pm 4.2)$ µJ.

**First THz seeded FEL experiments**

It should be noted that the THz radiation from the SASE FEL amplifier is only partially (transversely) coherent. To overcome this issue, a seeded THz FEL can be considered. The seeding concept assumes the introduction of a signal with an amplitude well above the intrinsic noise level and possibly a stable phase with a frequency in the bandwidth of the FEL amplifier. The seeding option has been widely considered at various facilities, from VUVs to X-ray FELs [42], [43], [44], [45]. The first experiments at PITZ with seeded THz FEL using temporally modulated photocathode laser pulses resulted in improved THz source performance in terms of peak pulse energy and shot-to-shot stability which was expected from the corresponding simulations [46].

Temporal modulation of the photocathode laser pulses was realized by means of special tuning of the Lyot filter installed in the regenerative amplifier of the laser system [47]. The temporal profiles of electron bunches with low charge (~10 pC, to minimize space charge effects) were measured using a transverse deflection system (TDS, see Figure 2). The smooth and modulated photocathode laser temporal distributions used for the SASE and seeding experiments, respectively, are shown in Figure 5a. The longitudinal beam momentum distributions in the low-energy dispersive arm (LEDA, see Figure 2) measured for a nominal bunch charge of 2 nC are shown in Figure 5b. The modulations from the photocathode laser are clearly imprinted in the longitudinal phase space distribution of the electron beam. While the modulations are longer than the FEL wavelength, strong space charge forces lead to nonlinear oscillations in the longitudinal phase space of the modulated electron bunch, resulting in an enhanced spectral density not only at the modulation frequency, but also in the higher frequency range, including the THz FEL resonant frequency. Thus, the startup radiation power in the seeding case is significantly higher than the shot noise level in the SASE case, resulting in a clear seeding effect. The corresponding gain curve measured for a 2 nC modulated electron beam is shown in Figure 5c in comparison with the SASE case (same blue curve as in Figure 3). The THz pulse energy optimization was performed at TD3 equipped with the band-pass filter. In addition to the increased THz pulse energy (33.5 μJ vs. 21.4 μJ), better pulse energy stability was also observed. This supports an experimental proof of seeding.

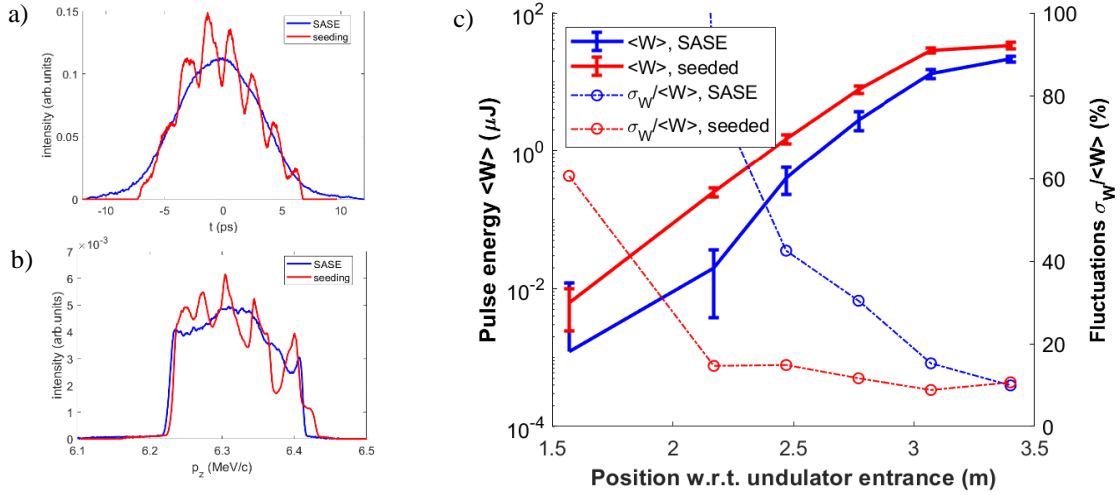

Figure 5: Seeded versus SASE FEL radiation. a) Measured temporal profiles of low charge electron bunches for smooth (SASE case) and modulated (seeded case) photocathode laser pulses. b) Corresponding beam longitudinal momentum distributions measured after the RF gun in a low-energy dispersive arm (LEDA, see Figure 2) for 2 nC bunches. c) Gain curves for 2 nC smooth (SASE case) and modulated (seeded case) electron bunches measured at TD3 (see Figure 2) with the band-pass filter. The pulse energy fluctuation rate is plotted on the right axis.

**Discussion**

A linear model of the FEL amplifier [29] can be applied to analyse the experimental data obtained. Using the linear 3D theory, and considering diffraction and space charge effects, one can obtain a radiation field gain $\Lambda$ (see *Appendix D*). Here, the radiation field amplitude along the undulator is assumed to be $E_x(z) \propto \exp[\Lambda \cdot z]$. Following this formalism and considering the case of 2 nC (blue curve in Figure 3) as a reference, one can obtain the gain parameter of the FEL amplifier $\Gamma = (0.24m)^{-1}$, given the measured peak current of ~125 A and beam energy of ~17 MeV. By applying these parameters, the FEL efficiency parameter $\rho \approx 0.01$ can be obtained. Applying the measured electron beam parameters to calculate the dimensionless parameters from [29] yields a beam diffraction parameter $B \approx 0.1$, which corresponds to a relatively thin beam. The space charge parameter $\hat{\Lambda}_p^2 \approx 0.9$ is quite large, as expected, so the space charge effect should impact the FEL field gain. An estimate of ~10 keV for the slice energy spread (the measured projected energy spread is ~70 keV) gives the energy spread parameter $\hat{\Lambda}_T^2 \approx 0.003$, which is small and can be neglected in the first approximation. On the contrary, the waveguide diffraction parameter $\Omega \approx 5.3$ is fairly large, which corresponds to the strong influence of the narrow vacuum chamber of the undulator onto the FEL field gain properties. Applying these dimensionless parameters to the eigenvalue problem for the field gain $\Lambda$ [29], [48] the gain (in units of $\Gamma$) can be calculated as a function of the frequency detuning $\Delta f = f - f_0$ from the resonance $f_0$; the results are shown in Figure 6a for four different cases: without and with space charge effect, and without and with the waveguide effect. As it is expected, the space charge effect leads to a decrease in the growth rate.

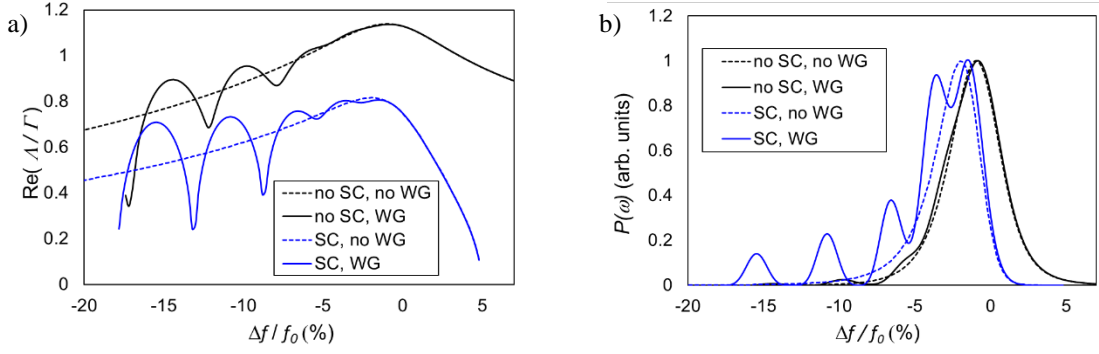

Figure 6: a) THz radiation field growth rate calculated using the waveguide version of the code FAST [48] for different cases as a function of the frequency detuning $\Delta f/f_0$: without (no SC) and with space charge (SC), without (no WG) and with the waveguide (WG) effect. b) Calculated radiation power spectra at the end of the high gain exponential regime for different cases.

The maximum value of the field growth rate from the linear theory with waveguide and space charge effects applied to the 2 nC case shows good agreement with the experimental gain curve (Figure 7). However, the beam wiggling amplitude in the middle of the undulator is estimated to be comparable to the RMS transverse beam size, which limits the applicability of the space charge model used in the linear model. The transverse waveguide modes calculated for negative frequency detuning values (local maxima of the solid curves in Figure 6a are mostly out of the used band-pass filter bandwidth. The double-peak structure revealed by the first FTIR measurements without band-pass filter (Figure 3c), but it should be mentioned that a similar structure is also obtained by FEL modeling without a waveguide effect (see simulations below).

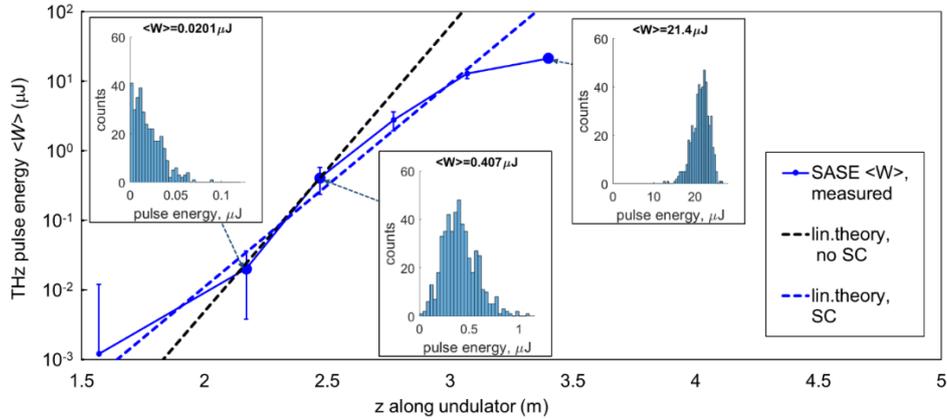

Figure 7: Linear theory fit to the measured THz pulse energy along the undulator using the maximum growth rates Re($\Lambda/\Gamma$) from the linear theory considering the waveguide effect without (no SC) and with space charge (SC) effect included in the field growth rate calculations (see Figure 6a). Insets: THz radiation pulse energy statistics measured for selected points of the 2 nC gain curve.

The measured fluctuations of the THz pulse energy were analyzed using a measurement statistic of 500 shots. Relevant radiation pulse energy distributions for points of the 2 nC gain curve (enlarged markers of the measured gain curve in Figure 7) are presented by histograms in the same figure. Numerical simulations of the experimental beam conditions were performed using the GENESIS code [35]. One of the main challenges is the proper modeling of the initial noise. The parameter space for the THz generation is very different from X-ray SASE FELs [1], [4]. In particular, the number of electrons per radiation wavelength is much higher for the THz case. If we compare XFEL radiation at 0.1 nm wavelength from electron beams with a peak current of 2 kA to THz generation at 100 μm from 200 A beams, the calculation of the number of electrons per radiation wavelength $N_\lambda$ yields $N_\lambda(THz)/N_\lambda(X-ray) \approx 10^5$. This ratio corresponds to very low shot noise level for the THz case, which makes the THz SASE simulations very sensitive to the initial conditions. On the other hand, the ratio of the electron beam length (~6 mm FWHM) to the radiation wavelength (~0.1 mm) is large, but not that extremely. The local bunching factor for time $t$ within the bunch can be calculated for all particles $N_t$ within the wavelength $\lambda$ around $t$ [49], [35]:

$$bf(t) = \frac{1}{N_t} \sum_{n \in N_t}^{c|t_n - t| \leq \lambda/2} e^{2\pi i \frac{ct_n}{\lambda}}. \qquad (2)$$

This calculation can be performed either for the real number of electrons in the beam or for the number of macroparticles used in the simulation, while modeling the required shot noise statistics. The local bunching factors used for FEL simulations are shown in Figure 8a along with the measured one, the beam current profiles are shown in Figure 8b. Assuming that the beam current profile is quasi-Gaussian, the contribution of the beam density slope within the radiation wavelength to the local bunching factor is much larger than the model shot (Schottky) noise level (dashed curve in Figure 8a).

In fact, there are other contributions to beam density fluctuations in intense electron beams [50], such as quantum effects during emission from the photocathode [51], space charge and intra-beam scattering processes during photoemission (known as Boersch effect [52]), beam transport [53], [54], and others. Start-to-end simulations for the experimental case were performed using the ASTRA [55] and GENESIS [35] codes, and these effects were not included in the simulations. The output particle distribution obtained from ASTRA simulations from the photocathode to the entrance to the undulator was used as an input for the THz FEL simulations with GENESIS. A smoothed current profile was used to generate electron bunch slices for the GENESIS simulation. The shot-to-shot noise in simulations is introduced by applying different base numbers for the Hammersley sequence generator. Analysis of the local bunching factor shows that there are two regions in the bunch (about half of the peak intensity) where the coherent part of the spontaneous emission is sufficiently large, which makes the considered case similar to the so-called self-amplified coherent spontaneous emission (SACSE [56]). The seeding case was modeled by appropriately modifying the bunching factor (red curve in Figure 8a). The simulated gain curves for the 2 nC reference case for SASE and seeded FEL are shown in Figure 8c. They show general agreement with the experimental results in terms of growth rate and relative seeding improvement compared to SASE, but the measured THz radiation pulse energy (despite the fact that it is necessary to scale up by a factor of ~2 to compensate for THz transport losses) is lower than in the ideal simulation and the FEL instability in simulations starts much earlier than in the experiment. One of the reasons for this discrepancy is probably the not fully optimized experimental case. This is evidenced by a significant increase in the energy of the radiation pulse measured at TD2 (see Figure 4) optimized at a later time. The late start of the FEL instability in the measured gain curves may be due to non-ideal electron beam overlap with radiation which is due to the lack of accurate beam diagnostics inside the undulator. Only in the middle of the undulator the beam is strongly focused (see Figure 10 in *Appendix B*), providing a strong initial signal for the FEL amplifier. On the other hand, the idealized simulation does not reproduce beam jitters, waveguide and wake-field effects due to the narrow vacuum chamber of the LCLS-I undulator. Also, space charge effects are limitedly modeled in the FEL simulations using the GENESIS code [35].

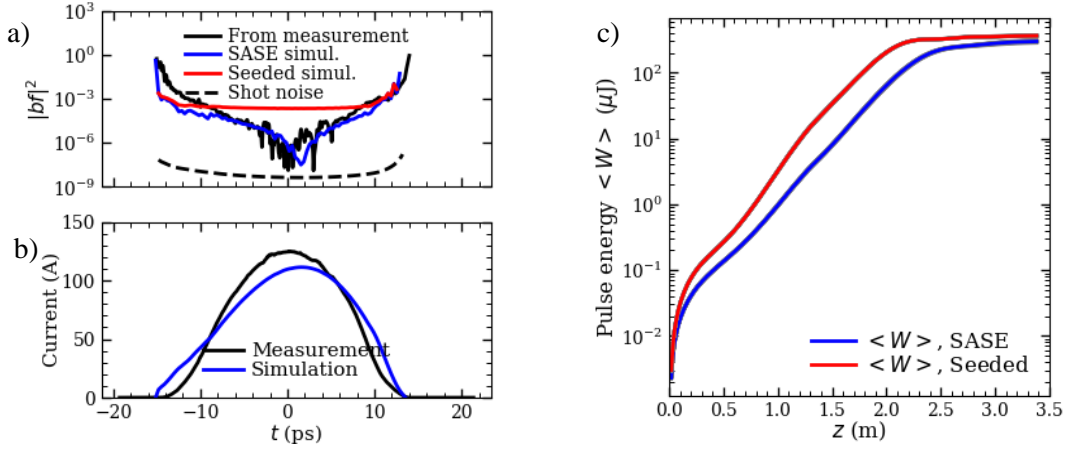

Figure 8: a) Bunching factors of 2 nC electron beam used for SASE and seeded FEL simulations. b) Simulated and measured beam current profiles. c) Gain curves simulated for THz SASE and seeded FEL for a 2 nC beam.

Simulated THz pulse temporal profiles with (5.9±0.1) ps RMS duration and corresponding spectra centered at (98.0±0.4) μm with (4.3±0.2) μm RMS bandwidth for the 2 nC SASE reference case at the undulator exit are shown in Figure 9a,b. The simulated spectrum in the middle of the undulator (z~2.5 m) is shown in Figure 9c. It has a double-peak structure similar to the experimental observations (see Figure 3 c). The simulated THz radiation pulse energy is (299.8±5.6) μJ. The simulated stability is significantly better than the measured one, which is a consequence of the limitations of the initial noise modeling on the one hand, and the jitters and beam fluctuations present in the experiment on the other (see *Appendix B* for some recorded beam jitters).

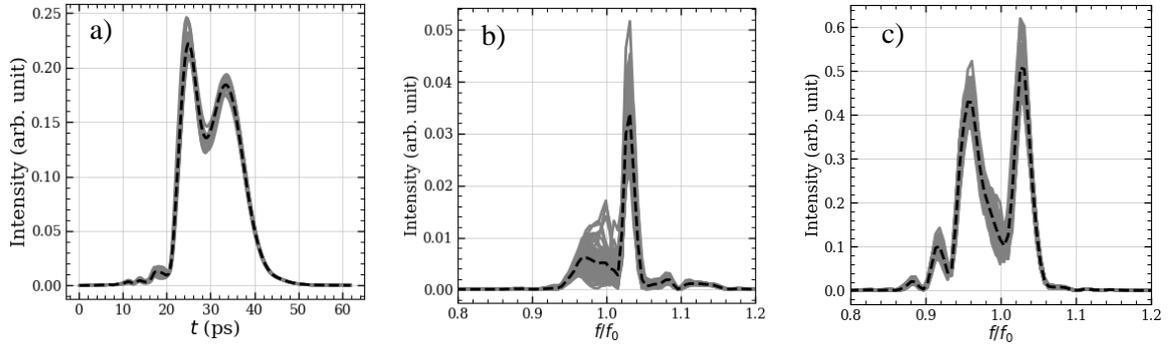

Figure 9: THz radiation pulse profile (a) and corresponding spectrum (b) at the end of the undulator simulated for the 2 nC SASE case. c) Radiation spectrum simulated in the middle of the undulator. The gray traces depict results of 100 simulations, the black curve is the average profile/spectrum respectively.

Further investigation of the parameter space in simulations resulted in a significant increase in the pulse energy of THz radiation when using temporal flattop photocathode laser pulses instead of Gaussian pulses (as used in the experiment). Combined with doubling the bunch charge, this yields an expected THz pulse energy of up to 1.5 mJ (see *Appendix E*). Such kind of flattop pulses are expected to be achievable at PITZ after the upgrade of the photocathode laser system which is currently underway at the facility.

## Conclusions and outlook

In summary, the reported experimental results already demonstrated the generation of narrowband THz radiation with a very high pulse energy of tens of microjoules by using the SASE FEL mechanism. Further improvements of the radiation properties are feasible. This proof-of-principle experiment with radiation at a central wavelength of ~100 micrometers was realized by combining an LCLS-I undulator and the PITZ high-brightness photo injector. Due to the full compatibility of the accelerator used for THz generation with the XFEL photoinjector, the presented high-power THz source is perfectly matched to MHz bunch repetition rates and pulse train structure of modern X-ray sources operated and under development at various facilities around the world, enabling efficient pump-probe experiments. Continuous tunability of such a THz source by adjusting the electron beam energy makes it a powerful tool for precise control of matter. It has been confirmed that the peak current, i.e., the bunch charge of several nC, is the most important pivot for the high performance of the THz source. The SASE mechanism involved assumes an electron bunch length much longer than the central radiation wavelength, allowing the high charge of the electron beam to be utilized without being limited to a short bunch length. Although in the considered case the radiation wavelength is much shorter than the electron bunch duration, it turned out that the contribution of the bunching factor arising from the beam current profile to the initial spontaneous emission is larger than the shot noise level. The corresponding modeling yields a more stable radiation than in the case of a standard SASE FEL driven only by shot noise. In addition, further manipulation of the bunching factor by temporal modulation of the photocathode laser pulse, carried out in the first experiments with seeded THz FEL, showed an improvement in the THz source performance. Another tool to increase the THz FEL pulse energy is photocathode pulse shaping, namely using a flattop temporal profile instead of a Gaussian one, which is expected to bring the radiation pulse energy to the millijoule level for 4 nC beams. This option will be investigated at PITZ in the future. In addition, THz FEL seeding was demonstrated by using temporally modulated photocathode laser pulses to drive the RF gun, which improved the performance in terms of pulse energy and shot-to-shot stability. To put the concept in perspective, accelerator-based THz FEL sources are on the horizon, fully supporting the required pulse train structure and providing millijoule level THz pulses with tunable central frequencies in the range of interest for pump-probe experiments at modern X-ray FELs.

# Acknowledgements

This work was supported by the European XFEL research and development program. We would like to thank the European XFEL Management Board for supporting these developments.

*Appendix A: LCLS-I Undulator*

The LCLS-I undulator module is a 3.4-m-long permanent magnet ($Nd_2Fe_{14}B$) planar hybrid structure with 113 periods of $\lambda_U$=30 mm and a magnetic gap of 6.8 mm [34]. The strong magnetic field ($B_{y,max} \sim 1.25$ T) also has a horizontal gradient, which was originally intended to fine-tune the undulator parameter. Magnetic measurements of the LCLS-I undulator module L143-112000-26 have been performed at DESY in Hamburg to check its status after its transportation from SLAC. The axial field profile is shown as inset in Figure 2. The measurements yielded the undulator parameter $K(x = 0) \approx 3.48$ and the horizontal gradient $\frac{dK}{dx}(x = 0) = 0.748 \text{ m}^{-1}$. The parameters of the undulator require an electron beam energy of ~17 MeV for a central radiation wavelength of ~100 µm. However, tracking the reference particle through the undulator using three-dimensional magnetic fields, which are accurately reconstructed from the measurements [57], revealed a significant influence of the horizontal field gradient on its trajectory. Indeed, the reference particle would hit the wall of the narrow vacuum chamber (11 mm width, 5 mm height) about halfway through the undulator. To mitigate this undesirable effect, special long air compensation coils were developed and implemented [57] to compensate the influence of the horizontal gradient of the undulator field on the beam trajectory, and keep the beam centroid close to the center of the vacuum chamber.

*Appendix B: Electron beam generation, transport and matching*

Electron beams with bunch charges of up to 3 nC were generated using 7 ps (FWHM) ultraviolet (UV) photocathode laser pulses by adjustment of the laser spot size at the cathode (∅ 2-3.5 mm) and the laser pulse energy. The generated high charge electron bunches are longer (~15-20 ps) than the laser pulses due to space charge effects. A beam mean momentum of ~6.3 MeV/c was measured after the gun with the gun operated at maximum acceleration phase, and the final beam momentum of 16.5-17 MeV/c was achieved by tuning the gradient and the phase of the booster cavity. The booster phase was chosen to be ~20 deg off-crest, which roughly corresponds to the minimum projected energy spread at the undulator entrance.

The strong magnetic field of the planar undulator results in very strong vertical focusing and thus requires a very thorough vertical matching of the space charge dominated beams into the LCLS-I undulator. At the same time, the horizontal matching is mainly determined by the space charge effect during transport inside the undulator. The last quadrupole triplet (see Figure 2) in front of the LCLS-I undulator is used for matching the Twiss parameters [58] of the electron beam at the undulator entrance. The design X- and Y- Twiss parameters are rather asymmetric and have relatively small acceptance. The matching of the high charge electron beam into the LCLS-I undulator requires the following transverse beam parameters: the rms beam sizes $\sigma_x \approx 1.3\ mm$ and $\sigma_y \approx 0.2\ mm$, the corresponding beta-functions $\beta_x \approx 10.9$ m and $\beta_y \approx 0.4$ m, which corresponds to a flat beam configuration at the undulator entrance [59], [60].

Smooth transport of the space charge dominated electron beam over the distance of more than 27 m from gun to undulator was realized using three quadrupole triplets. A typical 2 nC beam envelope (RMS beam sizes) simulated along the beamline is shown in Figure 10. The plot also shows experimental data: the RMS beam sizes measured at the available screen stations which are in adequate agreement with simulations. Corresponding electron beam distributions measured along the beamline are shown in Figure 2. The electron beam trajectory through the undulator (z=28.2 m to z=31.6 m) with a matching triplet (z~27.8 m) is also shown in Figure 10. The slightly increased discrepancy between experiment and modeling along the beamline is mainly due to imperfections such as inhomogeneities in the transverse beam distribution and wakefields that were not included in the simulations.

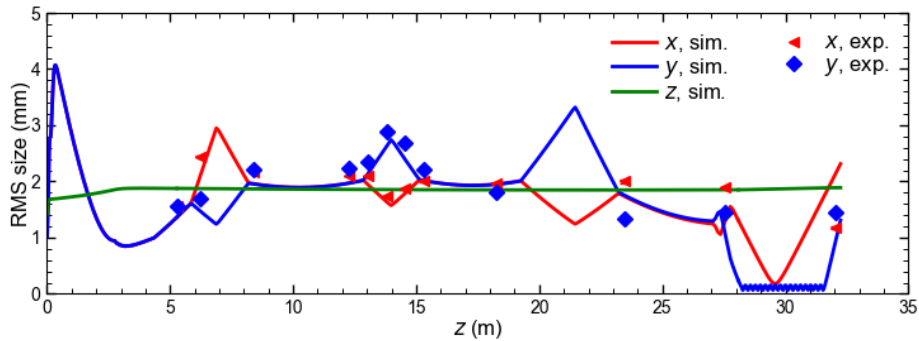

Figure 10: Beam envelope for 2 nC, 17 MeV/c electron beam. Solid curves: simulated RMS beam sizes; markers: transverse RMS beam sizes measured at the screen stations along the beamline. Statistical RMS errors on the measured beam sizes are obtained on the order of 1-2% and are hidden by the marker size. Some of the measured transverse electron beam distributions are shown in Figure 2.

The experimental electron beam parameters, including jitters, are summarized in Table 1 for the 2 nC reference case.

Table 1: Experimental electron beam parameters

| Parameter | Measurement location | | Unit | Mean value | RMS jitter |
|---|---|---|---|---|---|
| | Station | distance from photocathode | | | |
| Photocathode laser transverse distribution | CCD camera at the virtual cathode (VC2) | Equivalent to 0 m | | | |
| center <X> | | | mm | | 0.011 |
| center <Y> | | | mm | | 0.014 |
| Xrms | | | mm | 0.718 | 0.001 |
| Yrms | | | mm | 0.733 | 0.003 |
| Intensity (sum of pixels) | | | % | | 1.7 |
| Bunch charge | Faraday cup LOW.FC1 | 0.803 m | nC | 2.002 | 0.030 |
| Cathode quantum efficiency QE | Faraday cup LOW.FC1, photocathode laser power meter | 0.803 m | % | 6.7 | |
| Beam mean momentum from RF gun | Low energy dispersive arm (LEDA) | 1.7 m | MeV/v | 6.304 | 0.001 |
| Beam rms momentum spread from the RF gun | | | keV/c | 55.2 | 0.7 |
| Final beam mean momentum | After booster at High energy dispersive arm (HEDA) | 8.92 m | MeV/v | 17.086 | 0.003 |
| Final beam rms momentum spread | | | keV/c | 65.7 | 1.9 |
| Transverse normalized beam emittance | Emittance measurement system (EMSY1), slit scan | 5.27 m | | | |
| X-emittance | | | mm mrad | 5.41 (95%) | |
| Y-emittance | | | | 6.95 (90%) | |
| Beam current profile | Transverse deflecting Cavity (TDS) | 10.985 m | | | |
| Ipeak | | | A | 125.3 | 17.1 |
| Trms | | | ps | 5.4 | 0.8 |
| FWHM | | | ps | 19.0 | 1.2 |
| Beam upstream the undulator | High3.Scr1, last screen before the undulator | 27.558 m | | | |
| center <X> | | | mm | | 0.032 |
| center <Y> | | | mm | | 0.046 |
| Xrms | | | mm | 1.915 | 0.010 |
| Yrms | | | mm | 1.516 | 0.016 |
| Intensity (sum of pixels) | | | % | | 2.2 |

*Appendix C: **THz diagnostics***

The THz radiation measurement setup consists of three THz diagnostics stations (TD1-3), each of which is also equipped with scintillating (YAG) screens (High3.Scr1-3) to measure the transverse distribution of electron beam. The first station TD1 is located in front of the undulator and is used for electron beam matching as well as for measuring the electron beam bunching factor by means of coherent transition radiation (CTR). The diagnostics after the undulator is composed of two stations named TD2 and TD3 (Figure 11), a dipole magnet, and a beam dump. TD2 and TD3 with their respective High3.Scr2 and High3.Scr3 screens are located ~0.5 m and ~1.5 m downstream of the undulator exit, correspondingly. Each screen station has an in-vacuum gold-coated ellipsoidal mirror for transporting the photon beam from the beamline through a diamond window to a pyrodetector. The mirrors can be moved remotely in and out from the beamline using linear motorized stages. The diamond window has a clear aperture of 20 mm and has a flat transmission of >70% for a wavelength longer than 10 μm [39]. The mirror at TD2 has a diameter of 50.8 mm with a 5 mm to 8 mm conical hole that allows the electron beam to pass through. The electron beam is then bent to the beam dump using the dipole magnet.

Therefore, only the photon beam can travel to the mirror at TD3. This mirror also has a diameter of 50.8 mm but without a hole. Each screen station has a THz photon diagnostic setup installed at the diamond window's exit. A band-pass filter (BPF3.0-24 Tydex [61]) with maximum transmission of ~92%, centered at 102 μm and a bandwidth of ~12 μm (FWHM) was mounted on top of a cylindrical adapter in front of the pyrodetector at High3.Scr3. The THz radiation collection efficiency was estimated by light propagation in a free space using the LightPipes tool [62] to be 49% and 63% for TD2 and TD3, respectively [39].

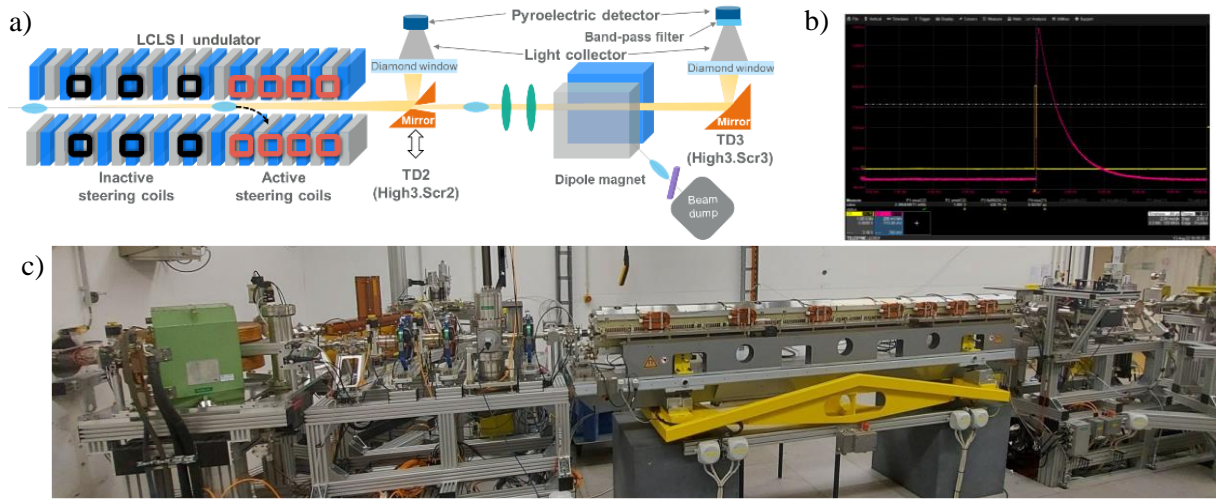

Figure 11: THz diagnostics setup. a) Schematic of the THz diagnostic stations and gain curve measurement setup. b) Typical pyroelectric oscilloscope signal (pink waveform). c) Photo of the THz beamline in the PITZ tunnel annex, electron beam direction is from left to right.

The pyrodetector used in the photon diagnostic setup is a room-temperature pyroelectric detector model THz10 manufactured by SLT Sensor- und Lasertechnik GmbH [63]. A signal voltage preamplifier is used to amplify the output signal from the pyrodetector with an adjustable gain of 10, $10^2$, $10^3$, and $10^4$. The thermal time constant (the relaxation time from the heated state to the ambient state) of this pyrodetector is about 50 ms. Therefore, the detector can measure pulse energies with a repetition rate of 10 Hz which is the normal repetition rate of the pulse trains at PITZ. Typically, a single pulse in a pulse train was used to measure the shot-by-shot statistics. Only when the signal-to-noise ratio was low the number of electron bunches in the pulse train was increased to measure the mean energy of the radiation pulse.

There is a set of short steering coils distributed along the undulator (Figure 11). They allow to kick the electron bunch horizontally away from the nominal trajectory in the undulator which is used to measure the THz pulse energy radiated till the kick location (active undulator length), which finally provides the gain curve. Indeed, the electron beam at this kick does not get into the beam dump and is lost, smearing on the side wall of the vacuum chamber. Due to moderate beam energy and rather large smearing spot, as expected, no damage to the hardware was observed so far. In the second part of the undulator the last four coils are more concentrated to provide more points of the gain curve near the location where saturation is expected to occur.

*Appendix D*: ***Linear model of the FEL amplifier***

Linear theory of the FEL amplifier with a planar undulator is based on self-consistent equations for radiation field and particle motion [29]. The radiation wavelength from the LCLS-I undulator by applying a beam energy of ~17 MeV can be calculated using $\lambda_{rad} = \frac{\lambda_U}{2\gamma^2}\left(1 + \frac{K^2}{2}\right) \approx 100 \mu m$, where the undulator parameters are taken from *Appendix A*. Parameters specifying the interaction of an electron beam with the electromagnetic field in a planar undulator are $Q$ and $A_{JJ}$, defined as:

$$Q = \frac{K^2}{4+2K^2} = 0.43, \tag{3}$$

$$A_{JJ} = J_0(Q) - J_1(Q) = 0.75, \tag{4}$$

where $J_n$ is the Bessel function of $n$-th order. The maximum angle of electron oscillations in the planar LCLS-I undulator $\theta_l = \frac{K}{\gamma} = 0.11\ rad$ (~6 deg), what corresponds to rather large wiggling amplitude. Thus, the longitudinal relativistic factor (as it defined by $\frac{1}{\gamma_l^2} = \frac{1}{\gamma^2} + \frac{\theta_l^2}{2}$) can be calculated to be $\gamma_l = 12.25$. The gain parameter of the FEL amplifier, $\Gamma$, defines the scale of the field gain for the 1D model and is equal to

$$\Gamma = \sqrt{\frac{I_{peak} A_{JJ}^2 2\pi^2 f_0^2 \theta_l^2}{I_A c^2 \gamma_l^2 \gamma}} = (0.24\ m)^{-1}, \tag{5}$$

where $f_0 = 2\gamma_l^2 c / \lambda_U$ is the resonance frequency of the amplified radiation field, $I_{peak}$ is the beam peak current, $I_A = \frac{mc^3}{e} \approx 17 kA$ is the Alfven current. The expected field gain length is of the order of 24 cm. Besides the gain factor $\Gamma$, some dimensionless parameters should be calculated in order to analyze various effects impacting the THz FEL amplifier operation mode. The 3D linear theory of the FEL amplifier includes also diffraction effects. The diffraction parameter $B = \frac{4\pi \Gamma \sigma_r^2 f_0}{c} \approx 0.1$ is close to the approximation of the thin beam and the diffraction effects will play a significant role in the THz FEL amplifier operation. To calculate the typical radial beam size $\sigma_r = \sqrt{2\langle\sigma_x\rangle\langle\sigma_y\rangle}$, needed to calculate the diffraction parameter $B$, averages along the undulator values ($28.2m \leq z \leq 31.6m$ in Figure 10) were used: $\langle\sigma_x\rangle \approx 0.75\ mm$ $\langle\sigma_y\rangle \approx 0.2\ mm$. Another important factor is expected to be the space charge effect. The space charge dimensionless parameter is $\hat{\Lambda}_p^2 = \frac{4c^2}{[\theta_l \sigma_r 2\pi f_0 A_{JJ}]^2} \approx 0.9$, thus the space charge effect cannot be neglected. An estimate on the efficiency parameter of an FEL amplifier theory $\rho = \frac{\gamma_l^2 \Gamma}{2\pi f_0/c} \simeq 0.0104$ is rather high compared with shorter wavelength FELs, like for the case of X-ray SASE FELs. The beam energy spread dimensionless parameter $\hat{\Lambda}_T^2 = \frac{\sigma_E^2}{[E_0 \rho]^2} = 0.003$ calculated from the expected slice energy spread of ~10 keV is also rather small and can be neglected in the first approximation. The three-dimensional linear theory of the FEL amplifier involves solving an eigenvalue problem for the radiation field [29]. In addition, the waveguide influence can be included in the form of waveguide mode decomposition [48]. The waveguide diffraction parameter $\Omega = \Gamma \cdot R_{eff}^2 \cdot \frac{2\pi f_0}{c} \approx 5.3$, where $R_{eff}^2 = \frac{a \cdot b}{\pi}$ is the square of the effective transverse size of the waveguide (undulator vacuum chamber with width $a$ and height $b$). The numerical solution of the above eigenvalue problem yields the radiation field gain $\Lambda$ ($E_x(z) \propto \exp[\Lambda \cdot z]$) as a function of the radiation frequency detuning $\Delta f$ from the resonant frequency $f_0$ (Figure 6).

*Appendix E*: *THz SASE FEL Simulations with 4nC beams*

In order to explore the parameter space towards the highest THz pulse energies that can be expected at the current PITZ THz beamline, the ultimate performance of the photoinjector was applied to THz SASE FEL simulations. Instead of a Gaussian temporal shape, flattop photocathode laser pulses with a FWHM of ~20 ps were used to generate 4 nC electron beams. Results of THz FEL simulations with GENESIS [35] for the center frequency of $f_0 = 3$ THz are shown in Figure 12.

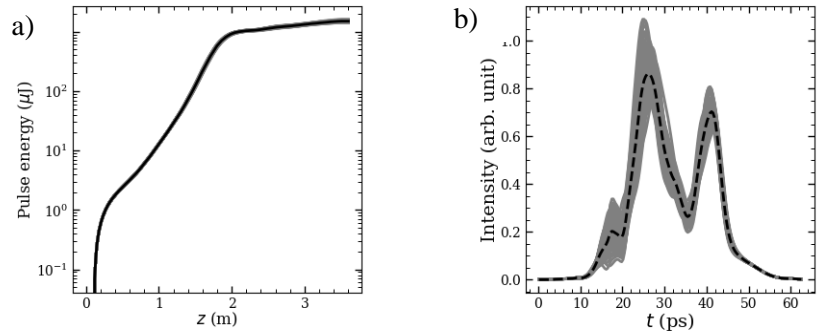

Figure 12: a) Gain curve simulated for the 4 nC SASE case using flattop photocathode laser pulses. THz radiation pulse profile (b) were calculated at the LCLS-I undulator exit. The gray traces depict results of 100 simulations, the black dashed curve is the average profile. The corresponding simulated spectrum is shown in Figure 1c.